\documentclass[print,review,12pt,authoryear]{elsarticle}

\usepackage[margin=1in]{geometry}

\usepackage{amssymb}
\usepackage{lscape}

 \usepackage{graphics}
 \usepackage{graphicx}
\usepackage{amssymb}
\usepackage{amsmath}
\usepackage{mathtools}
\usepackage{pgfplots}
\pgfplotsset{width=3cm,compat=newest}% <-- moves axis labels near ticklabels (respects tick label widths)
\usepackage{subcaption}
 \usepackage{lineno}
\usepackage{wrapfig}
\usepackage{lipsum}
\usepackage{textcmds}

\usepackage{array}
\usepackage{booktabs}
\usepackage{hyperref}
\usepackage{pgfplots}
\pgfplotsset{width=3cm,compat=newest}% <-- moves axis labels near ticklabels (respects tick label widths)
\usepackage{subcaption}
\usepackage{float}
\usepackage{graphicx}

\usepackage{graphicx}

\usepackage{soul, color}

\journal{ }

\begin{document}

\begin{frontmatter}

\title{Mode II fracture of an MMA adhesive layer: theory versus experiment}

%% use optional labels to link authors explicitly to addresses:
%% \author[label1,label2]{}
%% \address[label1]{}
%% \address[label2]{}
\author[cam_mic]{Sina Askarinejad}
\author[Imperial]{Emilio Mart\'{\i}nez-Pa\~neda}
\author[Burgos]{Ivan Cuesta}
\author[cam_mic]{Norman Fleck\corref{cor1}} 
\cortext[cor1]{Corresponding author}
\ead{naf1@cam.ac.uk}

\address[cam_mic]{Engineering Department, University of Cambridge, Cambridge, UK}%
\address[Imperial]{Department of Civil and Environmental Engineering, Imperial College London, London SW7 2AZ, UK}
\address[Burgos]{Universidad de Burgos, Escuela Politécnica Superior, 09006 Burgos, Spain}

\begin{abstract}
Thick adhesive layers have potential structural application in ship construction for the joining of a composite superstructure to a steel hull. The purpose of this study is to develop a mechanics model for the adhesive fracture of such lap joints under shear loading. Modified Thick-Adherend-Shear-Test (TAST) specimens made from a MMA-based adhesive and steel adherents are designed and fabricated. Crack initiation and growth of these joints is measured and monitored by Digital Image Correlation (DIC). An attempt is made to use a cohesive zone model to predict the magnitude of shear strain across the adhesive layer both at crack initiation and at peak load, and to predict the extent of crack growth as a function of shear strain across the adhesive layer. The ability of a cohesive zone model to predict several features of specimen failure is assessed for the case of an adhesive layer of high shear ductility.\\
\end{abstract}

\begin{keyword}
%% keywords here, in the form: keyword \sep keyword
Thick-adherend shear test \sep Fracture toughness  \sep Finite element modelling \sep Cohesive zone

%% PACS codes here, in the form: \PACS code \sep code

%% MSC codes here, in the form: \MSC code \sep code
%% or \MSC[2008] code \sep code (2000 is the default)

\end{keyword}

\end{frontmatter}

\newpage
\section{Introduction}
%\vspace{-2.5mm}

Mechanical joints made from bolts and rivets are commonly used in ship building. However, bonded joints offer significant benefits such as reduced weight, reduced through-life maintenance, and the reduction of the stress concentrations. A limitation in advancing this technology is the lack of confidence in the use of structural adhesive by the qualifying agencies. Despite the fact that adhesive joints are subjected to shear loading, studies on the mode II fracture of ductile adhesives have received limited attention in the literature \citep{yang2001,JAM2020}. Our aim is to investigate the effect of geometry and material parameters upon the shear strength of an adhesive joint for the case of an adhesive of high failure strain.\\

In the potential application of adhesive joints to large ships, it is envisaged that a thick adhesive layer (on the order of 10 mm) will be employed. Several studies have been carried out to investigated the effect of bond thickness on the fracture and failure of adhesive joints with an emphasis on thickness less than 1 mm \citep{mostovoy1971,kinloch1981,chai1986}. The sensitivity of fracture energy to the adhesive bond thickness is usually attributed to the interaction of plastic zone size and layer thickness \citep{hamoush1989,schmueser1990, Tvergaard1996}. In studies performed on mode I fracture of tough elastic-plastic adhesive/steel joints, the observed fracture changes from a single, flat cleavage plane at extremely small bond thicknesses (less than 0.4 mm) to river-like fracture surface marks indicative of ductile fracture for higher thicknesses \citep{yan2001, daghyani1995}. In those studies, the observed fracture energy for bond thickness in the range 1-10 mm is almost independent of thickness but is somewhat higher than that of thin (less than 1 mm) adhesive layers \citep{daghyani1995}. In contrast, much less is known about the failure of thick adhesive layers in shear. In our study, the mode II fracture toughness of an MMA adhesive layer, of thickness 3 mm to 13 mm, and sandwiched between steel substrates, is measured.\\

Linear Elastic Fracture Mechanics (LEFM) is commonly used to analyse the fracture of brittle adhesive joints \citep{williams1959, hutchinson2000}. Increasingly, toughened adhesives are used, and their non-linear elastic-plastic behaviour has generated the need for new approaches. There are two major developments in this field, one is the concept of the cohesive zone, as proposed by \citet{barenblatt1959} and \citet{dugdale1960}, and the other is the J-integral as proposed by \citet{Rice1968a}. The cohesive zone concept is based on a traction-separation law to describe the fracture behaviour, and the J-integral provides a convenient means of calculating the energy release rate at the crack tip. The use of the finite element method to model the failure of an adhesive zone by a cohesive zone idealisation is now established, particularly for mode I loading and for a linear elastic bulk response outside of the cohesive zone \citep{wei1998,williams2002,Feraren2004,ouyang2009}. In this method, toughness is not the only controlling parameter for fracture: the strength of the cohesive zone plays an important role too. Experiments are required to measure both the interfacial toughness and interfacial strength, and to determine the shape of the cohesive zone law. Among others, there exist two approaches for making use of a cohesive zone in modelling the failure of adhesive joints: (i) model the adherend and adhesive layer explicitly, and include a cohesive zone in order to represent the initiation and growth of a crack on a pre-determined plane \citep{kafkalidis2002, andersson2004}; and (ii) subsume the adhesive layer and the fracture process zone into the same row of cohesive elements \citep{madhusudhana2002}.\\

Few studies have focused on the failure of thick adhesive layers under macroscopic shear loading; in such cases failure can be accompanied by large shear strains in the adhesive layer and a non-linear response of the adhesive. The present study attempts to fill this gap by a combined experimental and modelling approach: the Thick-Adherend-Shear-Test (TAST) is performed on a methyl methacrylate (MMA) adhesive of high shear ductility. Insight is gained into the effect of pre-crack length and adhesive layer thickness upon the failure load. In order to model the observed interfacial fracture of these joints, a non-linear numerical analysis is used, incorporating a finite length adhesive layer and an appropriate cohesive zone model. The ability of cohesive zone modelling to predict the macroscopic failure shear strength and degree of sub-critical crack extension is thereby evaluated. \\

\section{Materials and Methods}

\subsection{Bulk properties of MMA adhesive} 
%\vspace{-2.5mm}

The adhesive is Methyl Methacrylate Adhesive (MMA), with trade name Scigrip SG300-40\footnote{Scigrip. Bentall Business Park, Glover Road, Washington NE37 3JD (UK)}, while the substrates of the TAST specimens are made from a low carbon steel. The uniaxial tension response of the adhesive was measured by casting dogbone specimens from the adhesive; the dogbones are of gauge length 33 mm, width 6 mm, and thickness 3 mm. Wedge grips were used to grip the ends of the dogbone specimens in the screw-driven test machine, and the axial strain over the gauge length was measured by a laser gauge along with reflecting tabs on the specimen. Cube-shaped specimens of side dimensions 6 mm were compressed between two smooth plates within the test machine to measure the uniaxial compressive response. 

\begin{figure}
\center
    \includegraphics[width=0.6\linewidth]{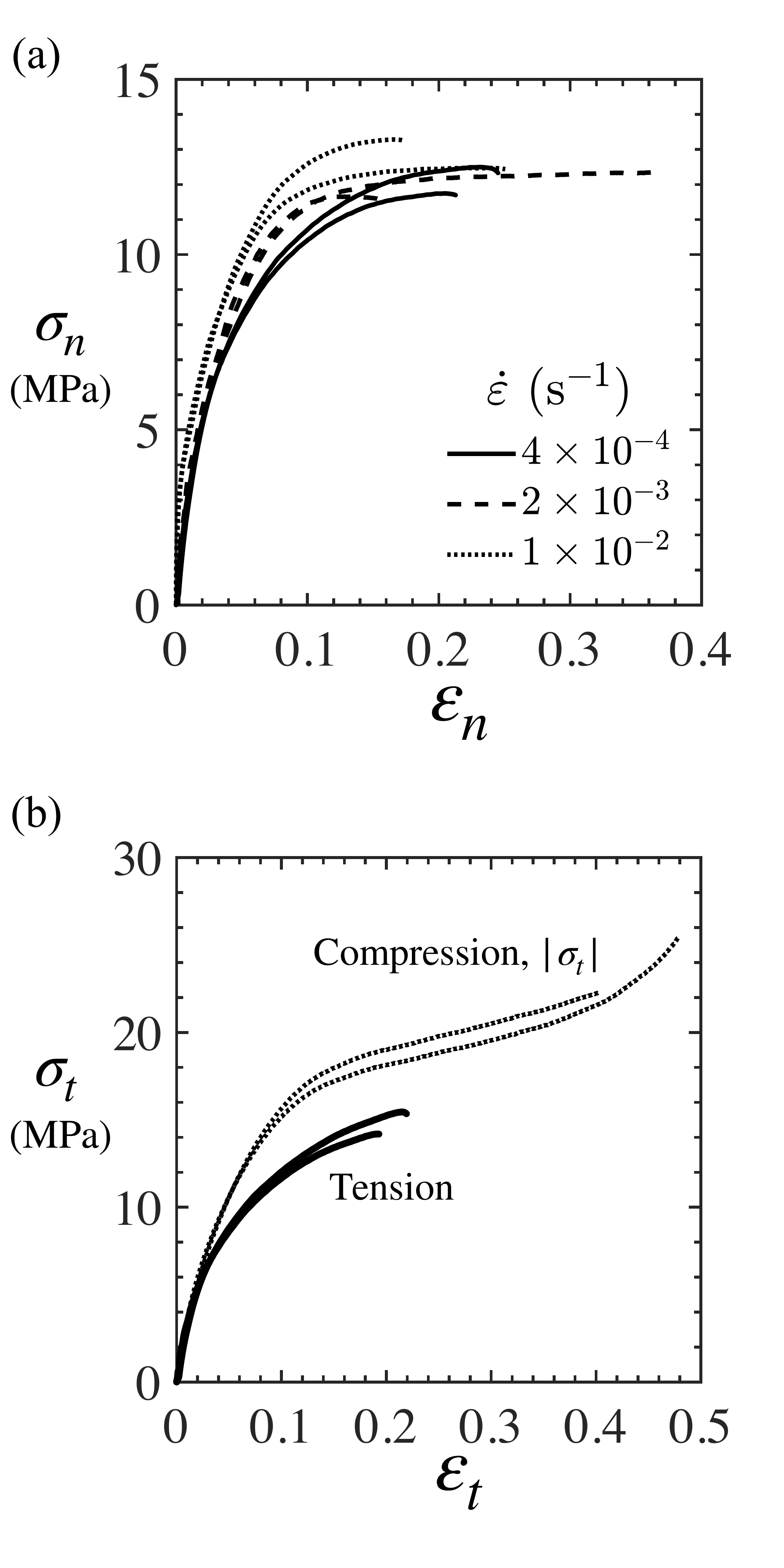}
\caption{(a) Nominal tensile response of Scigrip-300 adhesive in bulk form, for strain rates from $\dot{\varepsilon}=4\times10^{-4}$ s$^{-1}$ to $1\times10^{-2}$ s$^{-1}$, (b) True stress $\sigma_t$ versus true strain $\varepsilon_t$ curves in tension and compression, at $\dot{\varepsilon}=4\times10^{-4}$ s$^{-1}$.}
\label{PropSci}
\end{figure}

The uniaxial tensile stress versus strain response of the as-cured adhesive is given in Fig. \ref{PropSci}(a) for three values of nominal strain rate  $1 \times 10^{-2}$  s$^{-1}$, $2 \times 10^{-3}$  s$^{-1}$ and $4 \times 10^{-4}$ s$^{-1}$. Note that the nominal tensile yield strength increases from 6.2 MPa to 7.9 MPa by increasing the strain rate. The yield strengths are measured based on the 0.2 \% offset strain method. The Young's modulus of this adhesive varies from 262 $\pm$ 33 MPa to 312 $\pm$ 15 MPa within the range of strain rates considered. Moreover, the ultimate nominal tensile strength slightly increases from 11.6 MPa to 12.8 MPa. A wide range of scatter was observed for the nominal failure strain which does not significantly vary with strain rate. True stress versus strain responses under tension and compression with nominal strain rate of $4 \times 10^{-4}$ s$^{-1}$ are compared in Fig. \ref{PropSci}(b). The small elevation in compressive strength over tensile strength is consistent with the well-established pressure dependence of yields of polymers.

\subsection{Measured response of TAST adhesive joints}

\begin{figure}[h]
\center
    \includegraphics[width=1\linewidth]{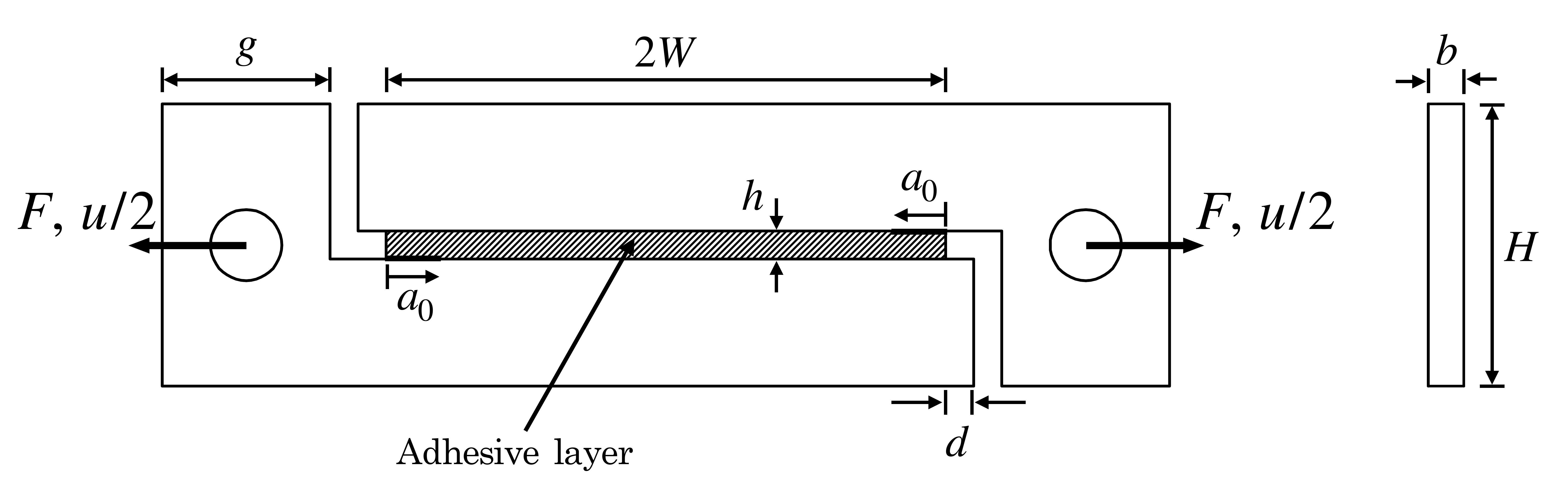}
\caption{Modified TAST specimen. $2W = 100$ mm, $h = 3-13$ mm, $H = 50.4$ mm, $g = 30$ mm, $a_0 = 0-20$ mm, $d = 5$ mm, $b = 6.35$ mm.}
\label{TAST}
\end{figure}

\begin{figure}
\center
    \includegraphics[width=0.7\linewidth]{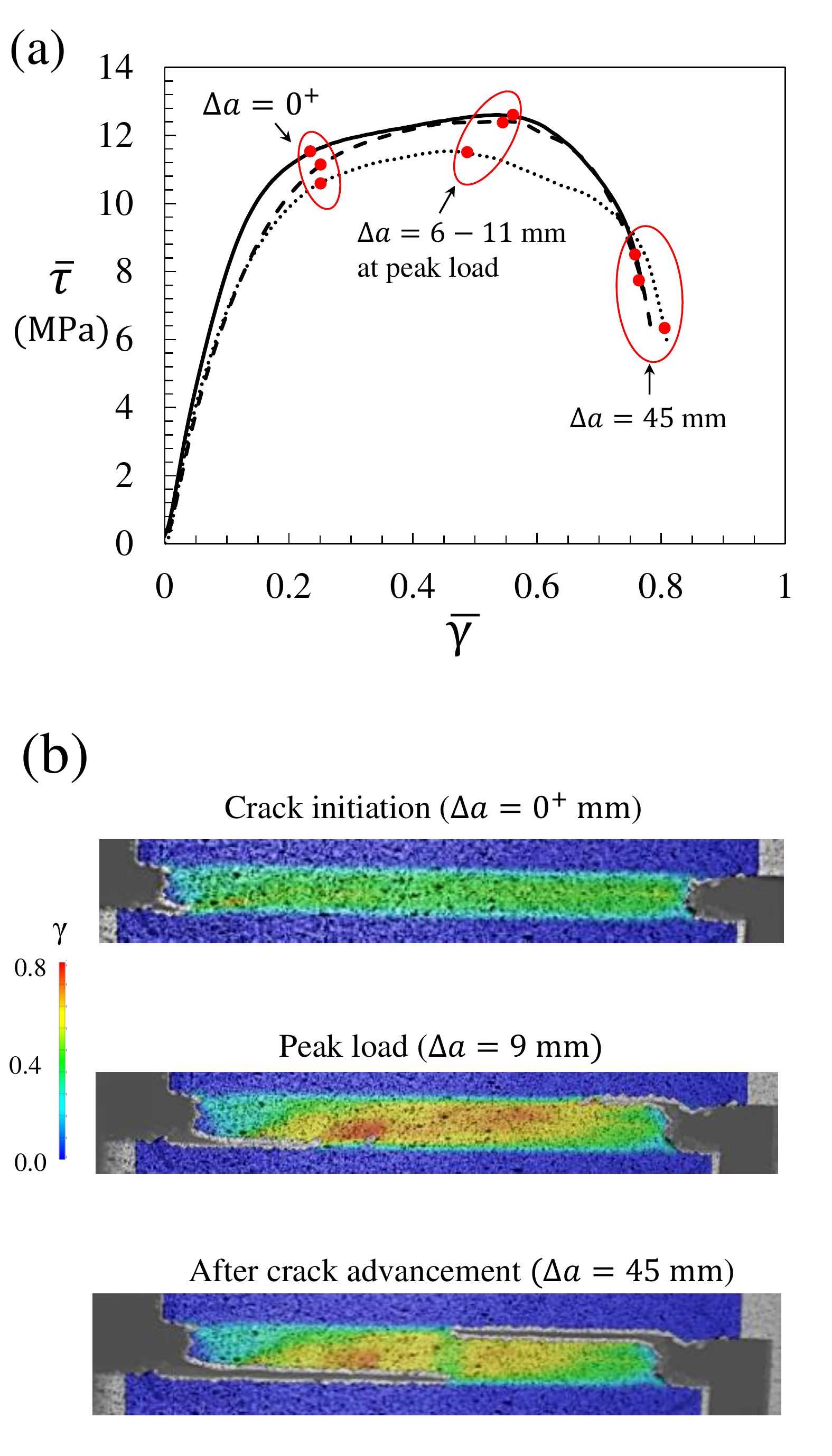}
\caption{(a) Typical force-displacement curve in a TAST test ($h =$ 8 mm, $a_0=$ 0, $\dot{\gamma} = 6 \times 10^{-4}$ s$^{-1}$, $\bar{\tau}=F/(2Wb)$, $\bar{\gamma}=u/h$), and (b) the corresponding DIC image of shear strain contour in each stage of deformation.}
\label{Rep}
\end{figure}

In order to evaluate the fracture response of MMA adhesive when sandwiched between steel adherends, a modified Thick-Adherend-Shear-Test (TAST) specimen was designed and manufactured (see Fig. \ref{TAST}).  The thickness $h$ of adhesive layer was varied from 3 mm to 13 mm, and the pre-crack length, $a_0$ was varied from 0 mm to 20 mm. In order to manufacture the modified TAST joints, the low carbon steel substrates were grit blasted and then degreased with acetone. The adhesive was applied (in accordance with the manufacturer's recommendations) using a manual applicator gun  with a static-mixing nozzle. The adhesive layer thickness, $h$, was adjusted by shims. All specimens were room temperature cured in ambient air for two weeks. Sharp pre-cracks were generated by razor tapping.
For each configuration, at least three specimens were tested. Crack extension was measured by using a high-resolution video-camera integrated in a Digital Image Correlation (DIC) system. 
The tests were conducted at room temperature (20$^\circ$ C) on the TAST specimens using a screw-driven test machine. The average shear strain rate within the adhesive layer was held fixed at $6 \times 10^{-4}$ s$^{-1}$ and, to accomplish this, the cross-head speed was varied in proportion to the adhesive layer thickness. The average shear stress $\bar{\tau}=F/(2Wb)$ versus average shear strain $\bar{\gamma}=u/h$ is plotted in Fig. \ref{Rep}(a) for the choice $a_0=0$ mm and $h=8$ mm. Repeat tests are included in the plot in order to indicate the degree of scatter. The DIC images of the side face of the specimen at the onset of crack initiation $\Delta a=0^+$, and for crack extensions $\Delta a=9$ mm and 45 mm, are shown in Fig. \ref{Rep}(b). The $\bar{\tau}$ versus $\bar{\gamma}$ response is initially linear, followed by a strain hardening regime before reaching a peak value and subsequent softening. Crack extension initiates ($\Delta a=0^+$) in the post-yield hardening regime, with peak stress accompanied by $\Delta a=9$ mm for the specimen reported in Fig. \ref{Rep}(b).

\subsection{Effect of adhesive thickness and pre-crack length on shear response of TAST specimens}

\begin{figure}
\center
    \includegraphics[width=0.75\linewidth]{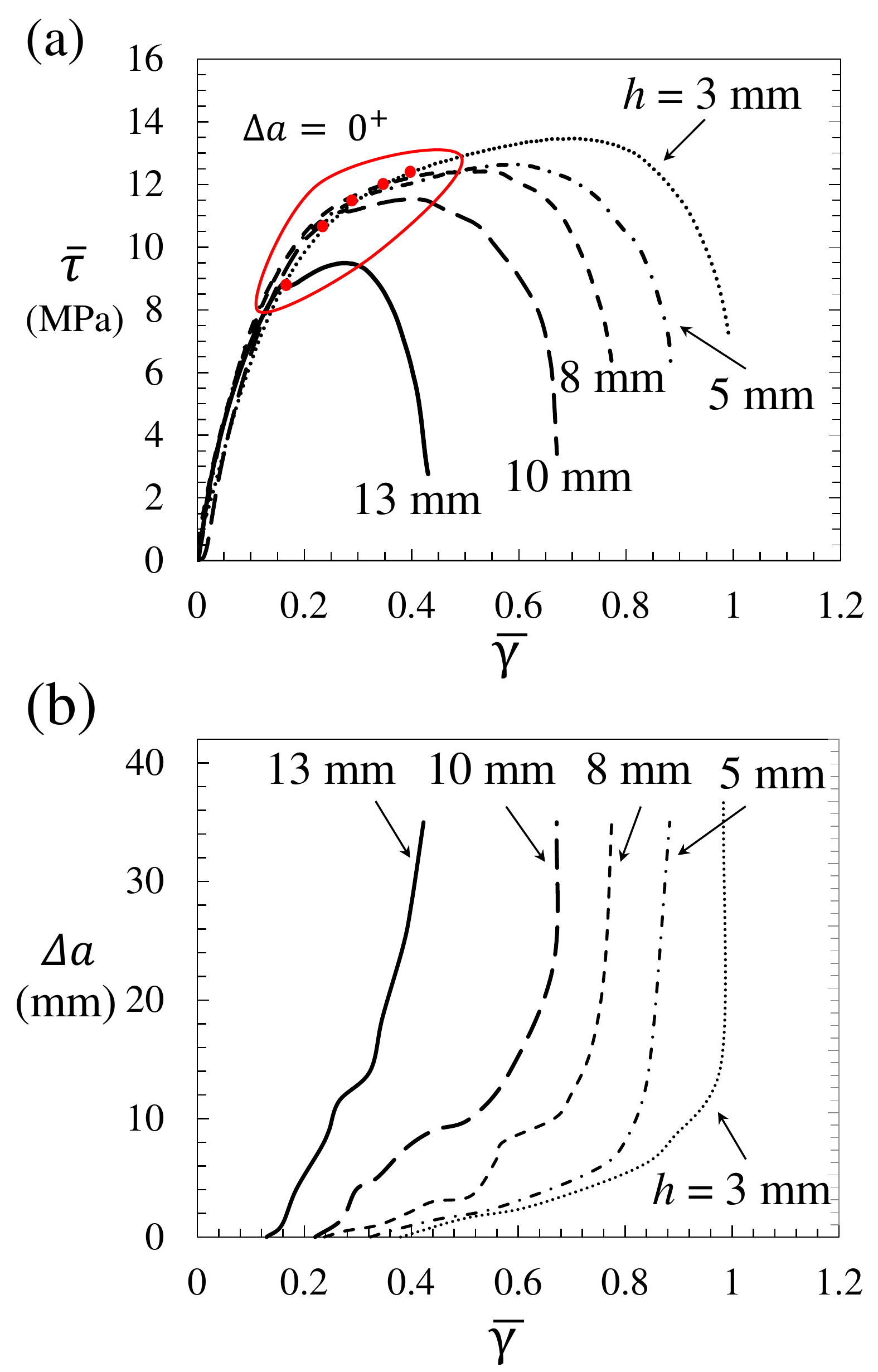}
\caption{Effect of adhesive thickness on samples with $a_0=$ 0 (no pre-crack). (a) average shear stress-shear strain across the adhesive layer curves (red marks on the curves shows $\Delta a =$ $0^+$ mm), (b) crack growth versus shear strain across the adhesive layer. }
\label{a0}
\end{figure}

The effect of adhesive layer thickness upon the shear stress $\bar{\tau}$ versus shear strain $\bar{\gamma}$ response and upon the crack extension response $\Delta a \left( \bar{\gamma} \right)$ is shown in Fig. \ref{a0} for the case of no pre-crack ($a_0 = 0$). The red dots indicate the onset of cracking, $\Delta a=0^+$. Both the shear strength (maximum $\bar{\tau}$) and the maximum achievable shear strain $\bar{\gamma}$ decrease with increasing layer thickness.\\

The value of $\bar{\gamma}$ corresponding to the onset of crack growth $\Delta a=0^+$ decreases with increasing layer thickness, see Fig. \ref{a0}(a). Note that the value of $\bar{\gamma} (\Delta a = 0^+ )$ much exceeds the yield strain (on the order of 0.1) in all tests. Some insight is given into the role of layer thickness $h$ on crack initiation in a related study \citep{Sina2020} on TAST tests for an elastic, brittle epoxy adhesive. In that study, the shear stress (and consequently shear strain) for crack initiation scaled as $h^{-1/3}$ due to the presence of a corner singularity at each end of the adhesive layer. In the present case a corner singularity persists but the adhesive is non-linear in response, and an analysis is more complex (and beyond the scope of the present study). Fig. \ref{a0}(b) shows the effect of adhesive layer thickness on crack extension of specimens with pre-crack length of $a_0 =$ 0 as a function of shear strain across the adhesive layer.\\

\begin{figure}
\center
    \includegraphics[width=0.75\linewidth]{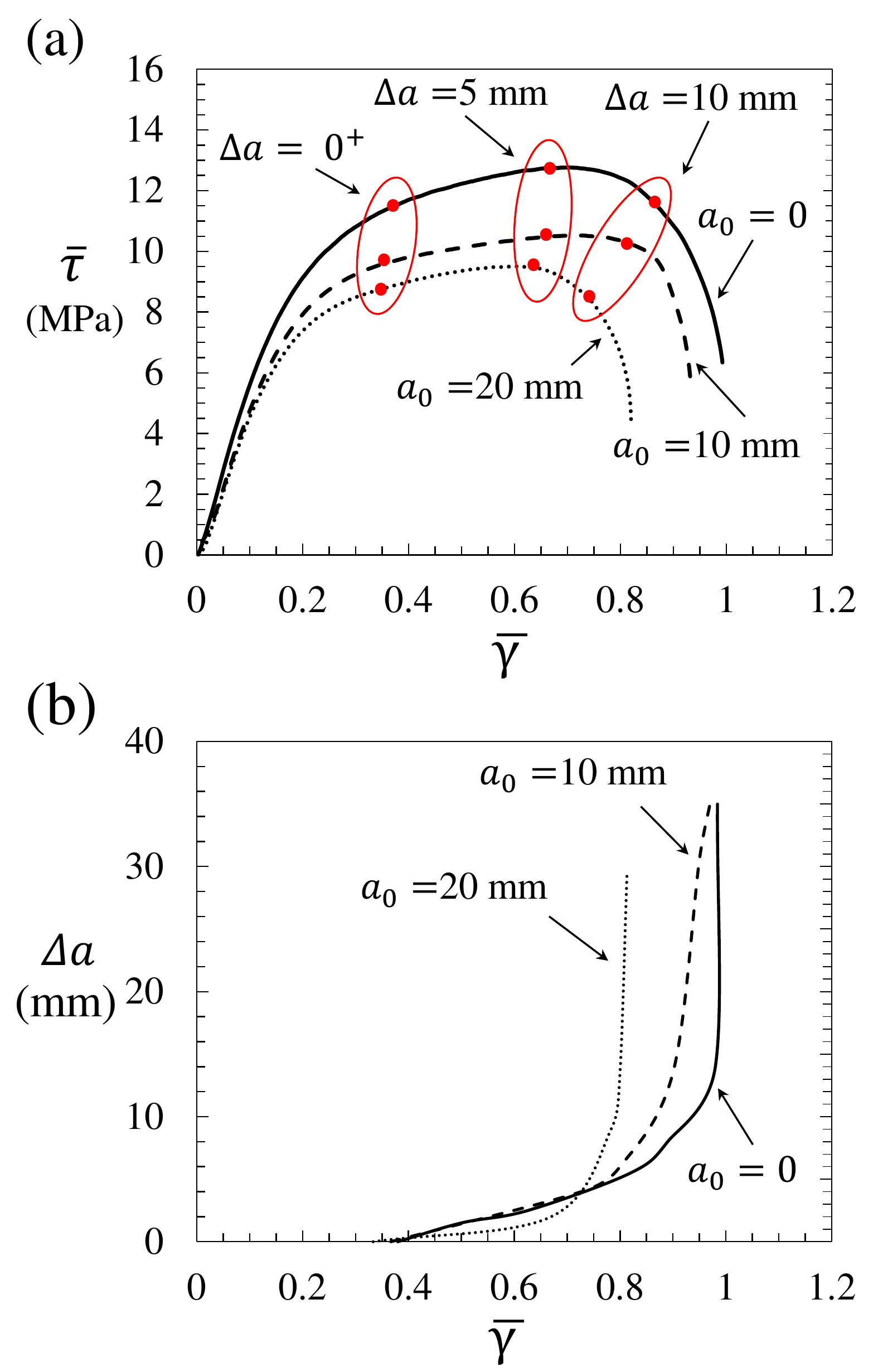}
\caption{Effect of pre-crack length on specimens with adhesive layer thickness of 3 mm. (a) average shear stress as a function of shear strain across the adhesive layer (marks on the curves shows $ \Delta  =$ 0$^+$, 5, and 10 mm), (b) crack growth versus shear strain across the adhesive layer for these cases. }
\label{h3}
\end{figure}

In order to evaluate the effect of pre-crack length upon the fracture response, specimens of adhesive layer thickness $h =$ 3 mm, and pre-crack length $a_0 =$ 0, 10, and 20 mm, are compared in Fig. \ref{h3}. The average shear stress versus shear strain responses across the adhesive layer are compared in Fig. \ref{h3}(a). Red dots on the $\bar{\tau}$ versus $\bar{\gamma}$ curves indicate the degree of crack extension ($\Delta{a}=$ 1, 5, and 10 mm). Both the shear strength and the level of shear strain across the layer at failure decrease as the pre-crack length $a_0$ increases. Fig. \ref{h3}(b) shows the effect of pre-crack length on crack extension of specimens with adhesive layer thickness of $h =$ 3 mm as a function of shear strain across the adhesive layer. The degree of crack extension $\Delta a \left( \bar{\gamma} \right)$ up to a shear strain of $\bar{\gamma}=0.7$ is almost independent of the value of $a_0$. At $\bar{\gamma}>0.7$ the rate of crack extension $\Delta a$ with respect to $\bar{\gamma}$ increases with increasing value of $a_0$.\\

\subsection{Critical crack sliding displacement}

\begin{figure}
\center
    \includegraphics[width=1\linewidth]{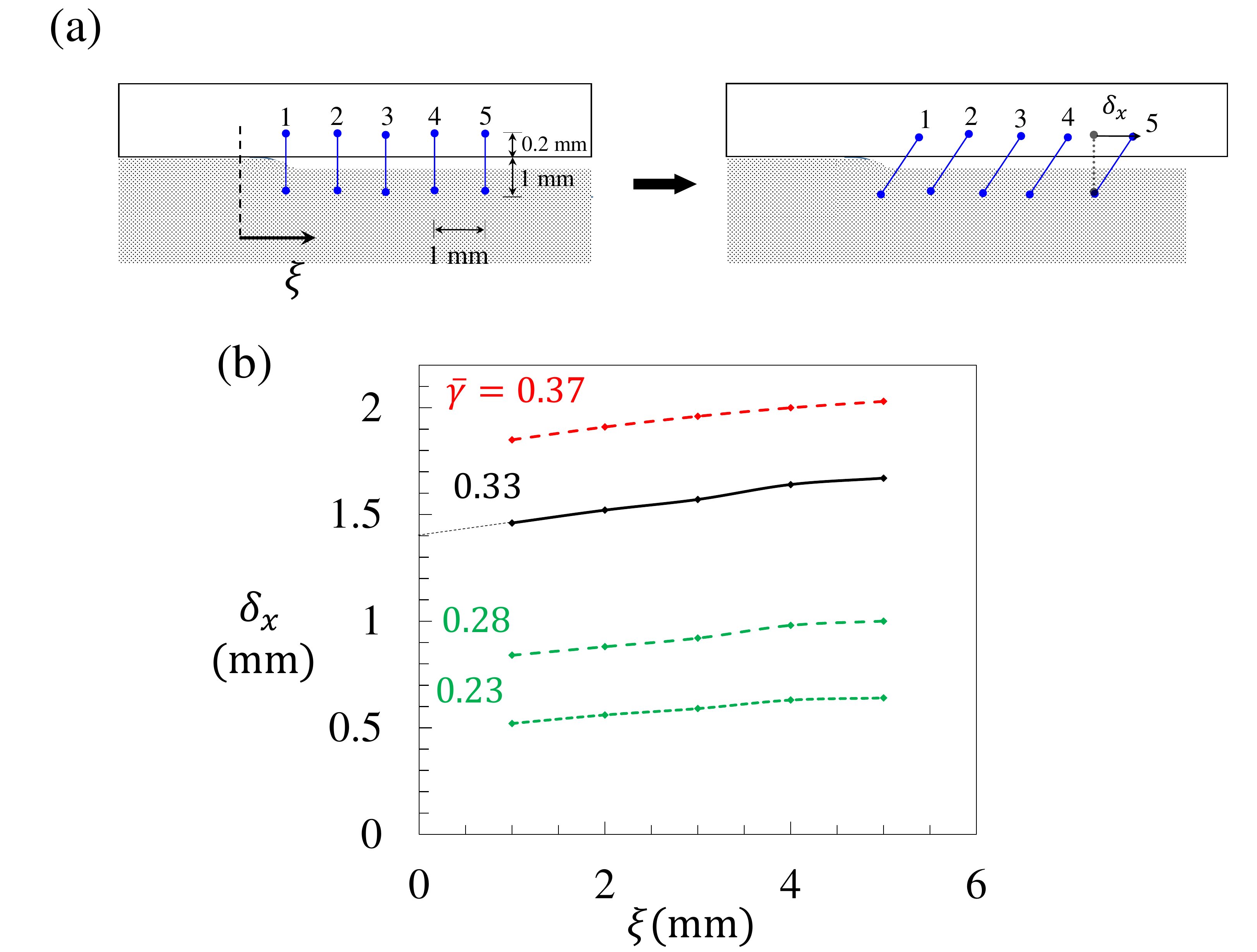}
\caption{(a) Method used to obtain sliding displacement jump $\delta_x$ at the crack tip prior to crack growth. (b) Sliding: displacement profile for the choice $h =$ 8 mm, $a_0 =$ 20 mm. Note, in (a), the finite opening of the pre-crack along the crack flanks; it is due to the saw-cut in manufacture. The tip of the pre-crack is sharp, due to razor tapping.}
\label{deltac}
\end{figure}

The magnitude of the critical shear displacement at the crack tip of a pre-crack, at the onset of crack growth, is measured by means of DIC. As shown in Fig. \ref{deltac}(a), five digital gauges are placed behind the crack tip. The spacing of the gauges is 1 mm, and the first one is 1 mm behind the crack tip. Figure \ref{deltac}(b) shows the mode II shear displacement versus distance from crack tip, for a representative case ($h =$ 8 mm and $a_0 =$ 20 mm), as measured by the digital gauges. For this specimen, crack advance initiates ($\Delta a=$ 0$^+$) at  $\bar{\gamma} =$ 0.33 and the jump in crack tip shear displacement is 1.4 mm. This measurement was repeated for specimens with $h$ in the range of 3 mm to 13 mm and $a_0=0$ and 20 mm. Thereby, an average value of $\delta_c=$ 1.3 mm was obtained. 

\section{Finite element modelling}
\label{sec:FEMmodel}

\begin{figure}
\center
    \includegraphics[width=0.7\linewidth]{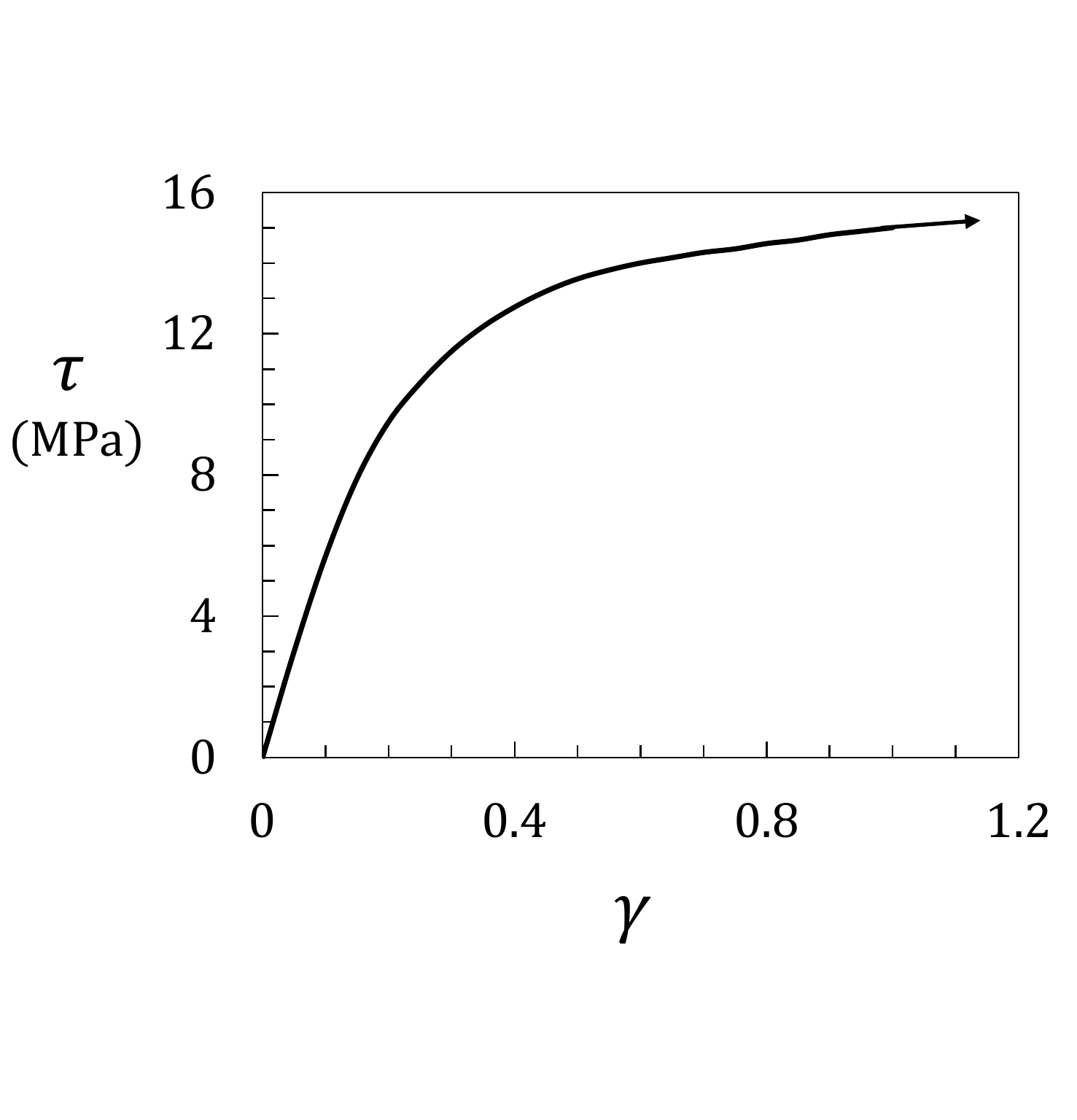}
\caption{The measured shear stress-strain for the adhesive ($\dot{u}/h$ = 6 $\times$ $10^{-4}$ s$^{-1}$) from a specimen of height $h=3$ mm, $a_0=0$.}
\label{FE}
\end{figure}

The steel has a sufficiently high yield strength ($\sigma_Y=300$ MPa) compared to the adhesive that it behaves in a linear elastic manner, with a Young's modulus $E=220$ GPa and a Poisson ratio $\nu=0.3$ \citep{yan2014}. The material response of the adhesive is modelled using J2 flow theory. Isotropic hardening is assumed and the material stress-strain curve is obtained from averaging six replicate shear experiments of $h=3$ mm, $a_0=0$. The resulting shear stress-strain curve is shown in Fig. \ref{FE}, with linear extrapolation assumed for $\gamma>1.0$. \\

\begin{figure}
\center
    \includegraphics[width=0.7\linewidth]{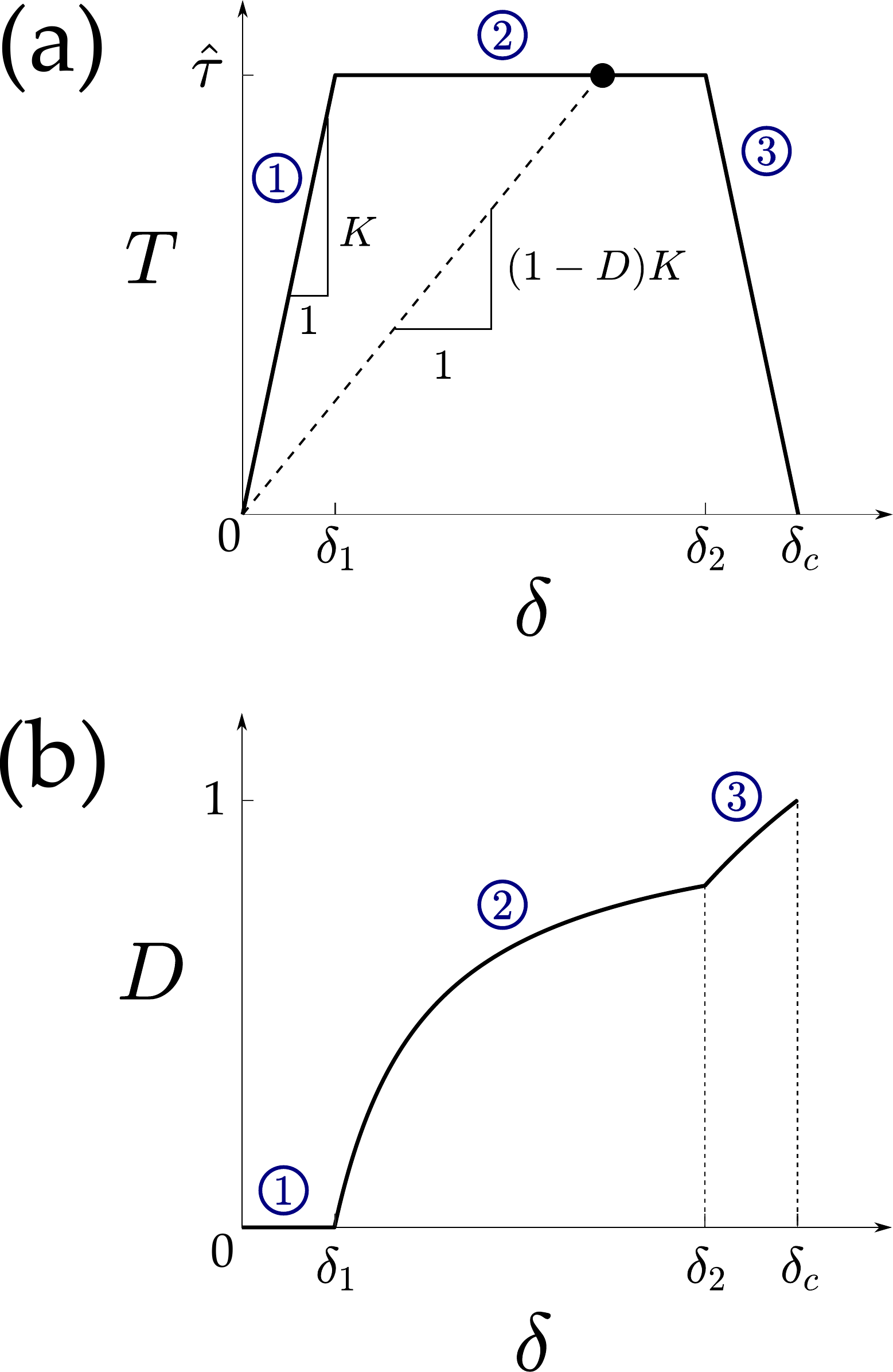}
\caption{Shear traction-separation response: (a) trapezoidal traction-separation law, with  secant modulus $(1-D)K$; and (b) damage evolution versus separation. In this study: $\delta_{1} =$ 0.05 $\delta_{c}$ and $\delta_{2} =$ 0.95 $\delta_{c}$.} %($\hat{\tau}$ = 13 MPa, , $\delta_{c}$ = 1.3 mm),}
\label{fig:FE2}
\end{figure}

Damage and failure of the adhesive layer is modelled via cohesive zone elements placed along each interface between the adhesive and the substrates. We limit our simulation of crack growth to samples with long pre-cracks $a_0>10$ mm as these specimens exhibit mode II fracture, circumventing the complexities associated with cohesive zone modelling of crack initiation from the interface corner under changing mode-mix \citep{tvergaard1993, camanho2003, Sina2020}. Thus, for our purposes, it is adequate to consider the progressive failure of a shear cohesive zone (see Fig. \ref{fig:FE2}). Failure of the adhesive/substrate interface is idealised by an assumed shear traction $T$ versus separation law $\delta$, of initial slope $K$. Following \citet{tvergaard1992}, a trapezoidal shape is assumed for the $T (\delta)$ relation, as characterised by three values of the separation ($\delta_1$, $\delta_2$, $\delta_c$) and a shear cohesive strength $\hat{\tau}$. The work of separation per unit area is defined as
\begin{equation}
    \Gamma_0 = \frac{1}{2} \hat{\tau} \left( \delta_c + \delta_2 - \delta_1 \right)
\end{equation}

We hold fixed the ratios $\delta_1/\delta_c=0.05$ and $\delta_2/\delta_c=0.95$, and thereby treat $\delta_c$ and $\hat{\tau}$ as the two primary parameters that define the cohesive zone law. A scalar damage variable $0 \leq D \leq 1$ is defined in terms of the secant modulus $T/\delta=(1-D)K$. As shown in Fig. \ref{fig:FE2}, the traction-separation law can be divided in three regions such that,

\begin{equation}
 T =
  \begin{cases}
    K \delta       & \quad \text{if } \delta \leq \delta_1 \\
     \hat{\tau}      & \quad \text{if } \delta_1 \leq \delta \leq \delta_2 \\
    \hat{\tau} \left( 1 - \frac{\delta - \delta_2}{\delta_c - \delta_2} \right) & \quad \text{if } \delta_2 \leq \delta \leq \delta_c
  \end{cases} \, \, \, \, \, \, ; \, \, \, \, \, \, \, \, \,    D =
  \begin{cases}
   0       & \quad \text{if } \delta \leq \delta_1 \\
     1 - \frac{\hat{\tau}}{K \delta}      & \quad \text{if } \delta_1 \leq \delta \leq \delta_2 \\
    \hat{\tau} \left( 1 - \frac{\delta - \delta_2}{\delta_c - \delta_2} \right) & \quad \text{if } \delta_2 \leq \delta \leq \delta_c
  \end{cases}
\end{equation} .

\noindent The damage $D$ versus separation $\delta$ response is shown in Fig. \ref{fig:FE2}(b). The cohesive zone model is implemented in the commercial finite element package ABAQUS by making use of cohesive surfaces. The trapezoidal law is introduced in ABAQUS by stating the damage $D$ versus separation $\delta$ characteristic in tabular form. The model is discretised with linear quadrilateral elements and a mesh sensitivity study is conducted to ensure that the fracture process zone is adequately resolved \citep{EFM2017}. 

\subsection{Calibration of the cohesive zone law}

\begin{figure}
\center
    \includegraphics[width=0.7\linewidth]{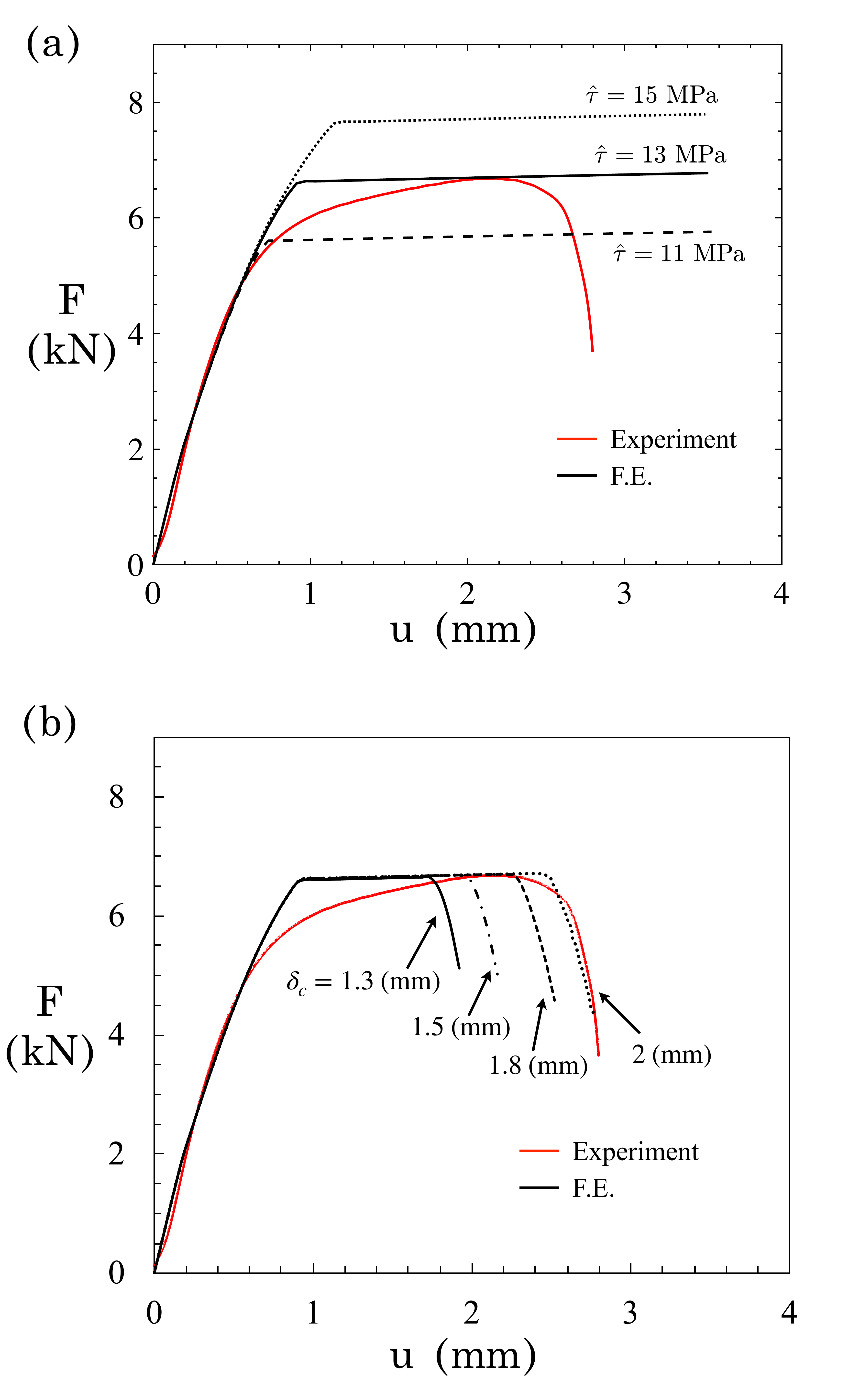}
\caption{(a) Effect of assumed value for $\hat{\tau}$ on the obtained force-displacement curve, (b) Effect of $\delta_c$ on load-displacement curve obtained from finite element simulations compared to the reference experiment ($h =$ 3 mm, $a_0 =$ 10 mm). Note: From DIC, $\delta_{c}=$ 1 - 1.5 mm.}
\label{Cal}
\end{figure}

The value of cohesive zone strength $\hat{\tau}=13$ MPa is chosen by matching the predicted shear strength of the TAST specimen to the observed value for the choice $h=$3 mm and $a_0=$10 mm, see Fig. \ref{Cal}(a). In the simulations an unbounded toughness is assumed such that $\delta_c =\infty$. It remains to choose a value of $\delta_c$ for the cohesive zone law. This is achieved by matching predictions of force $F$ versus displacement $u$ to the measured response in Fig. \ref{Cal}(b) for the TAST specimen of geometry $h=3$ mm and $a_0=10$ mm. Selected values of $\delta_c$ in the range of $\delta_c=1.3$ mm to $\delta_c=2$ mm are used in the predicitions. Best agreement is obtained for the choice $\delta_c=2.0$ mm. Recall that a direct measurement of the critical sliding displacement for the initiation of crack advance is $\delta_c=1.3$ mm from Fig. \ref{deltac}(b) and the related discussion.

\subsection{Accuracy of the cohesive zone model}

\begin{figure}
\center
    \includegraphics[width=1\linewidth]{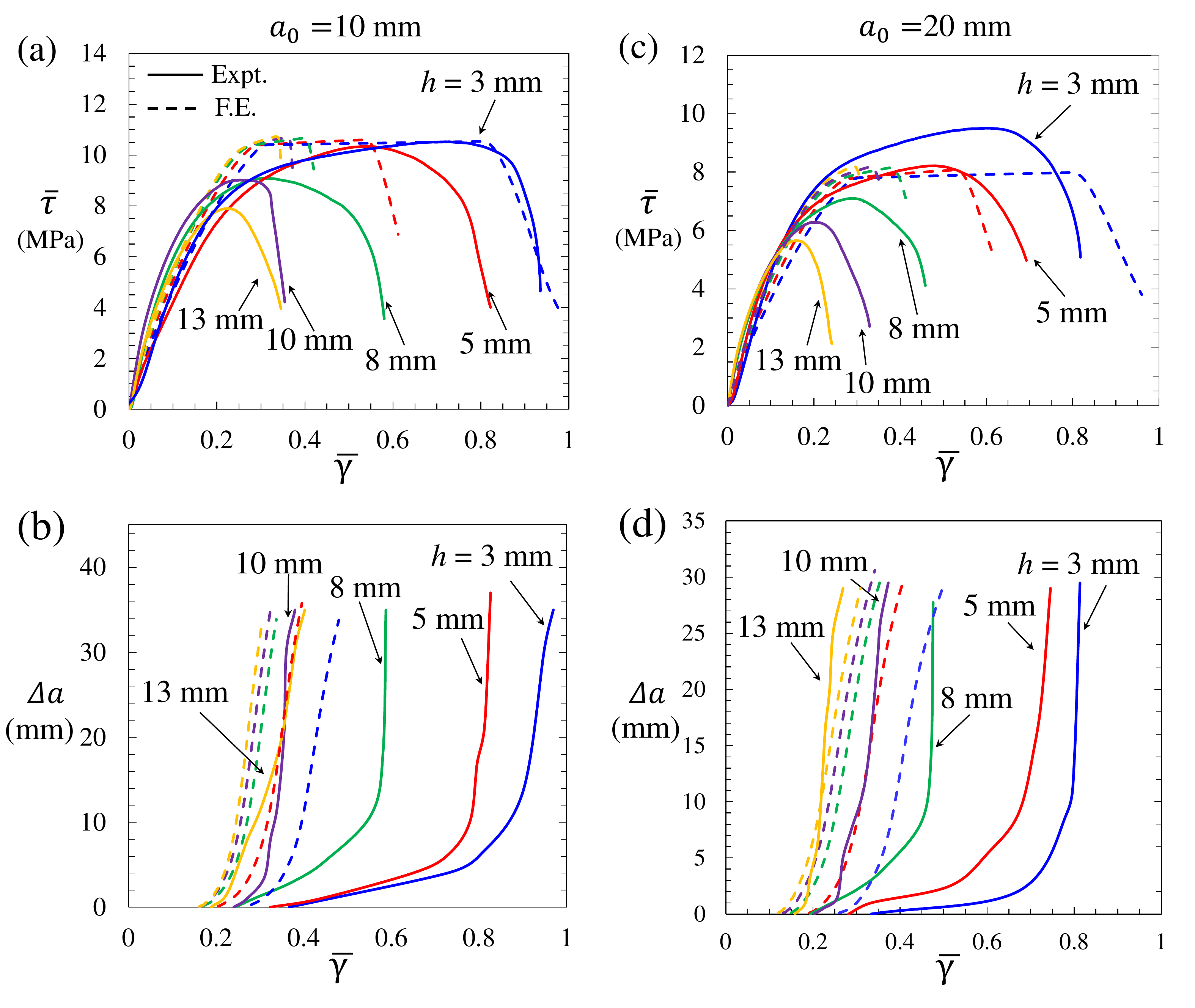}
\caption{Experiment vs simulations for (a) and (b) $a_0 =$ 10 mm; (c) and (d) $a_0 =$ 20 mm, assuming $\delta_c =$ 2.0 mm. Parts (a) and (c) give the average shear stress versus shear strain $\bar{\gamma}$ across the adhesive layer, from experiments and simulations. Parts (b) and (d) show crack extension $\Delta a$ as a function of $\bar{\gamma}$ across the adhesive layer.}
\label{caseDeltaC2}
\end{figure}

\begin{figure}
\center
    \includegraphics[width=1\linewidth]{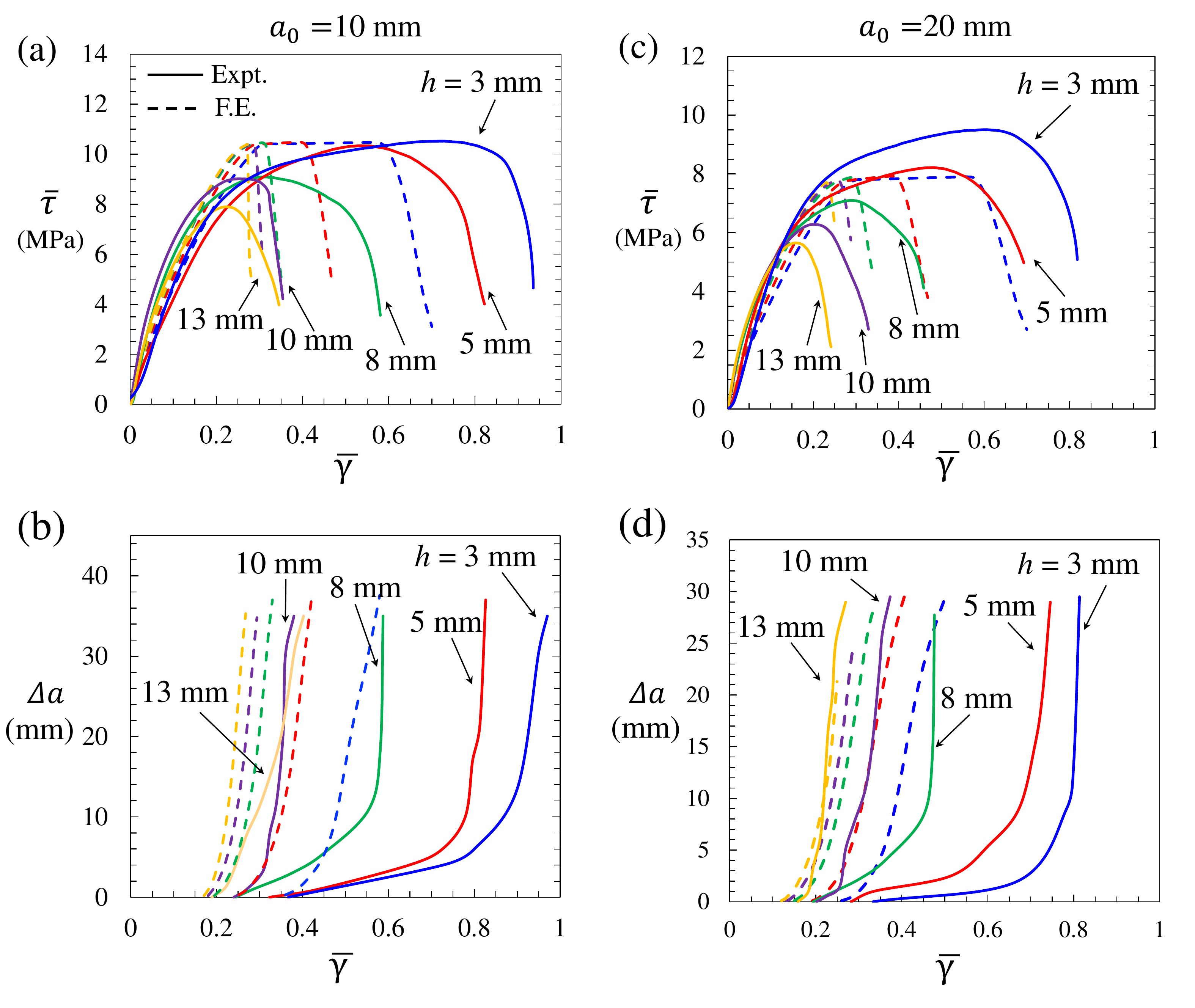}
\caption{Experiment vs simulations for (a) and (b) $a_0 =$ 10 mm; (c) and (d) $a_0 =$ 20 mm, assuming $\delta_c =$ 1.3 mm. Parts (a) and (c) give the average shear stress versus shear strain $\bar{\gamma}$ across the adhesive layer, from experiments and simulations. Parts (b) and (d) show crack extension $\Delta a$ as a function of $\bar{\gamma}$ across the adhesive layer.}
\label{caseDeltaC13}
\end{figure}

\begin{figure}
\center
    \includegraphics[width=0.7\linewidth]{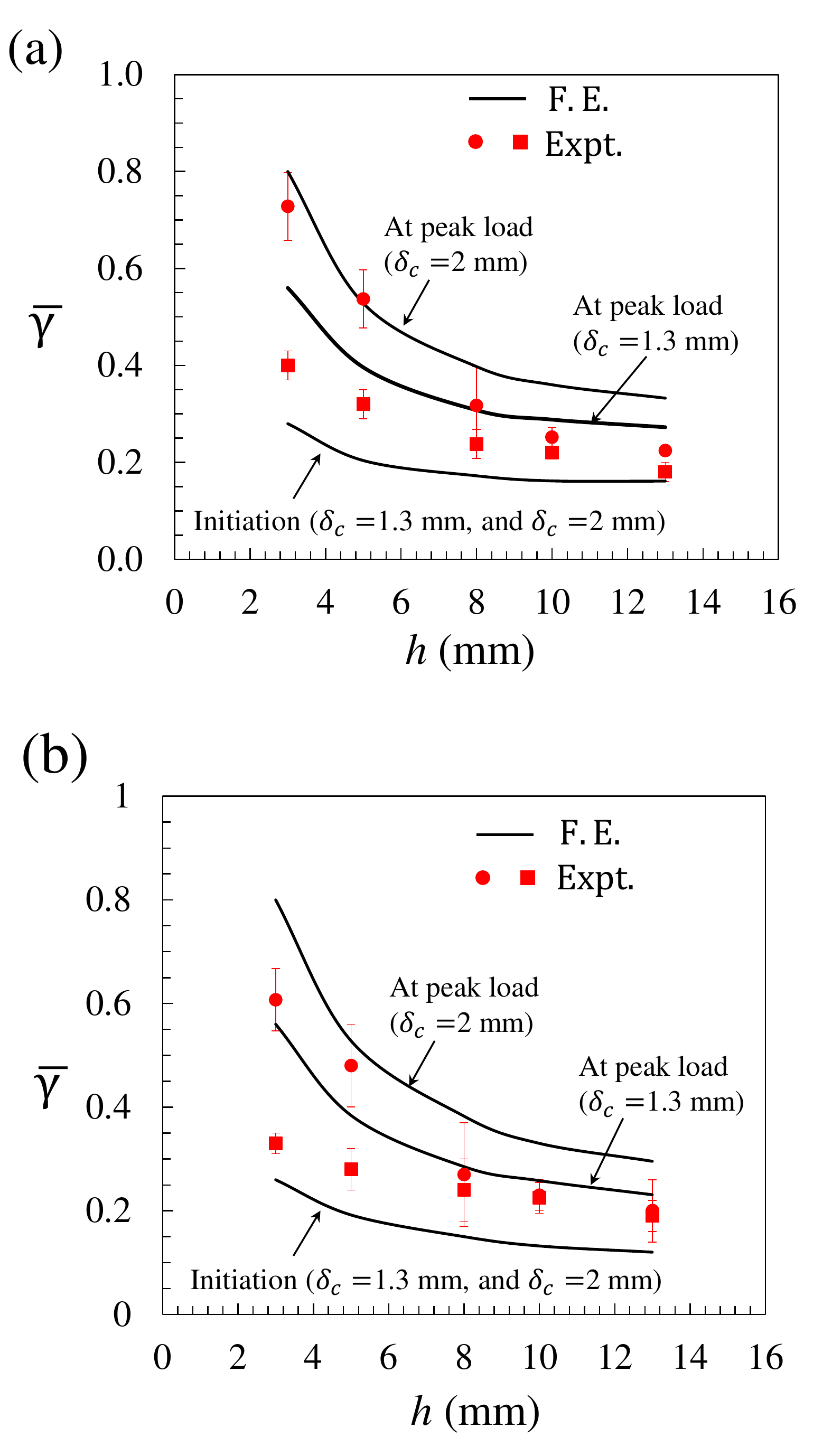}
\caption{Shear strain $\bar{\gamma}$ across the adhesive layer at crack initiation and peak load versus adhesive layer thickness $h$ for (a) $a_0 =$ 10 mm, and (b) $a_0 =$ 20 mm.}
\label{Fig11}
\end{figure}

Recall the definitions of average shear stress $\bar{\tau}=F/(2Wb)$, shear strain $\bar{\gamma}=u/h$ and crack extension $\Delta a$. Finite element predictions of $\bar{\tau} \left( \bar{\gamma} \right)$ and $\Delta a \left( \bar{\gamma} \right)$ are compared with the measured responses of TAST specimens in Fig. \ref{caseDeltaC2} for the choice $\delta_c=2.0$ mm, and in Fig. \ref{caseDeltaC13} for $\delta_c=1.3$ mm. In parts (a) and (b) of each figure we consider the tests with $a_0=10$ mm, whereas in parts (c) and (d) the tests are for $a_0=20$ mm. The cohesive zone model gives an accurate prediction of $\bar{\tau} \left( \bar{\gamma} \right)$ for the case $a_0=10$ mm, $h=3$ mm and $\delta_c=2.0$ mm; this is not surprising as the values of $\delta_c$ (and $\hat{\tau}$) for the cohesive zone model were chosen for this choice of geometry. However, the cohesive zone model (with $\delta_c=2.0$ mm) is less accurate in the prediction of $\Delta a$ $\left( \bar{\gamma} \right)$ for the choice $a_0=10$ mm and $h=3$ mm, see Fig. \ref{caseDeltaC2}(b). In broad terms, the cohesive zone model, in its present form, has a limited ability to predict the sensitivity of the $\bar{\tau} \left( \bar{\gamma} \right)$ curves and the $\Delta a \left( \bar{\gamma} \right)$ responses over a range in values of $h$ and $a_0$, for either choice of $\delta_c$. It is instructive to summarise the TAST results and associated predictions by plotting in Fig. \ref{Fig11} the shear strain $\bar{\gamma}$ for the initiation of crack growth, and the value of $\bar{\gamma}$ at peak $\bar{\tau}$, each as a function of $h$. In the experiments, at small values of  $h$ ($<6$ mm), the value of $\bar{\gamma}$ to attain peak load is almost double that required to initiate a crack. In contrast, for $h \geq10$ mm, crack initiation occurs at peak load. The cohesive zone calculations correctly predict that the value of $\bar{\gamma}$ for crack initiation, and to achieve peak load, both decrease with increasing $h$; however, the cohesive zone calculations predict significantly lower values of $\bar{\gamma}$ for crack initiation than the observed values.

\subsection{Alternative modelling approaches}

Our comparison of finite element predictions and TAST measurements suggest that it is challenging to use a cohesive zone model to predict failure of a ductile adhesive of limited work hardening capacity and high value of strain prior to cracking. The relatively flat stress versus strain curve implies that triggering of the cohesive zone is sensitive to the choice of cohesive strength $\hat{\tau}$. Modified approaches within the realm of cohesive zone modelling include strain-dependent cohesive zone models \citep{Tvergaard1996b} or triaxility-dependent traction-separation laws \citep{Anvari2006,Banerjee2009}. Alternatively, models based on continuum damage mechanics concepts can be employed.\\

Damage mechanics models typically have a damage initiation criterion, which can  be stress, strain, or energy-based, and a damage evolution law \citep{Kachanov1986,Simo1987,Chaboche1988a}. The material stiffness is degraded progressively according to the evolution of the damage variable $D$; $D$ is usually a scalar, but can be tensorial when damage is anisotropic in nature. The main drawback of conventional damage mechanics models is the strong mesh dependency due to the loss of ellipticity of the governing equations \citep{Needleman1988,Peerlings2002}, although numerical techniques have been developed to alleviate such mesh sensitivity \citep{Oliver1989}. The problem can also be regularised using higher order models, also termed non-local, involving gradient quantities and one or more associated length scales \citep{Geers1998,Bazant2003}. These models have been widely used in recent years, both for capturing size effects and for avoiding a pathologically localised post-peak response \citep{DeBorst1999,Engelen2006,JMPS2019}. Closely related to non-local damage models are phase field fracture models, where the phase field (damage) variable evolves on the basis of Griffith's energy balance \citep{Francfort1998,Bourdin2000,TAFM2020}. Thus, phase field models provide a link between damage mechanics and discrete cracking.\\

The consideration of a strain-based or energy-based cracking criterion appears to be better suited than a traction-based cohesive zone law for capturing the failure of adhesives that undergo large irreversible strains with limited hardening. The need to model toughness and the attendant material length scale suggests the use of non-local models involving spatial gradients or the integration of relevant quantities over a finite volume. However, is not yet clear which class of models can capture, for a single set of parameters, the multiple fracture features (peak load, fracture process zone size, critical crack opening) that accompany the failure of adhesive joints.

\section{Conclusions}
%\vspace{-2.5mm}

A combined numerical and experimental study of the Mode II fracture of a ductile adhesive joint is reported. A modified TAST setup, with a thick layer of MMA adhesive, is used in order to explore the effect of adhesive layer thickness on the initiation and growth of an interfacial crack between MMA adhesive and steel substrates. Regardless of pre-crack length, the adhesive layer deforms by a large shear strain of 0.18 to 0.35 (depending on the adhesive layer thickness) prior to crack initiation.\\

The ability of a cohesive zone model, with a trapezoidal shear cohesive zone law, to predict failure of an adhesive lap joint is evaluated. After suitable calibration, the shear cohesive zone is capable of predicting the fracture strength and shear strain at peak load of a TAST specimen for a limited range of layer thickness $h$. The model is less accurate in predicting the details of crack extension as a function of shear strain. The study suggests that a cohesive zone model, based on a traction-displacement law, requires careful calibration when applied to highly ductile adhesive layers that display limited strain hardening and a large plastic strain prior to crack initiation. 

\section{Acknowledgements}

This project has received funding from the Interreg 2 Seas programme 2014-2020 co-funded by the European Regional Development Fund under subsidy contract No. 03-051. The authors would also like to acknowledge financial support from the European Research Council in the form of an Advance Grant (MULTILAT, 669764).

\newpage 

\bibliographystyle{elsarticle-harv}
\bibliography{References-Scigrip}

%\newpage
%\listoffigures

\newpage

\newpage

\newpage

\newpage

\newpage

\newpage

\newpage

\newpage

\newpage

\newpage

\newpage

\newpage

\end{document}